\documentclass[
               ,final
               ]{aipproc}

\layoutstyle{6x9}

\usepackage{epsfig}  
\usepackage{amssymb} 
\usepackage{amsmath}     

\begin{document}  

\title{Perturbative Quantum Analysis and Classical Limit 
       of the Electron Scattering by a Solenoidal Magnetic Field}
       


\author{Gabriela Murguía}{
  address = {Departamento de Física, Facultad de Ciencias, UNAM.
         Apartado postal 70-542, 04510, M\'exico, D.F. M\'exico.}
}
\author{Matías Moreno}{
  address = {Instituto de Física, UNAM.
         Apartado postal 20-364, 01000, M\'exico, D.F. M\'exico.}
}
\author{Manuel Torres}{
 address = {Instituto de Física, UNAM.
         Apartado postal 20-364, 01000, M\'exico, D.F. M\'exico.}
}
\date{\today}

\begin{abstract}  

  A well known example in quantum electrodynamics (QED) shows that
  Coulomb scattering of unpolarized electrons, calculated to lowest
  order in perturbation theory, yields a results that exactly
  coincides (in the non-relativistic limit) with the Rutherford
  formula.  We examine an analogous example, the classical and
  perturbative quantum scattering of an electron by a magnetic field
  confined in an infinite solenoid of finite radius. The results
  obtained for the classical and the quantum differential cross
  sections display marked differences.  While this may not be a
  complete surprise, one should expect to recover the classical
  expression by applying the classical limit to the quantum result.
  This turn not to be the case. Surprisingly enough, it is shown that
  the classical result can not be recuperated even if higher order
  corrections are included.  To recover the classic correspondence of
  the quantum scattering problem a suitable non-perturbative
  methodology should be applied.
    
\end{abstract}  

\maketitle

\section{Introduction}   
\label{sec:Introduction}

As it is widely known, the scattering of unpolarized electrons by the
Coulomb potential exactly coincides with the classical Rutherford
formula, if one consider the lowest order in perturbation theory and
the non-relativistic regime.  In this paper, we examine an analogous
example: the scattering of an electron of momentum $p$ by a magnetic
field in a long solenoid of fixed flux $\Phi$ and finite radius $R$,
looking both at the classical and quantum regimes.

In the zero radius limit, the differential cross
section (DCS) is given by  the famous Aharonov-Bohm (AB)
result~\cite{AB}:
\begin{equation}  
 \left.{\frac{d\sigma}{d\theta}}\right|_{AB}=  
        \hbar \, \frac{\sin^2{\left(e\Phi/2\hbar c\right)}}  
        {2 \pi p \sin^2{(\theta/2)}}.
\label{eq:secc_dif_AB}  
\end{equation}
This is a purely quantum effect, because in the 
$\hbar \rightarrow 0$ limit  the expression cancels.

For a finite value $R$ of the solenoid radius, the classical cross
section will have a definite non-vanishing value, as far as the
electron can penetrate inside the solenoid. We shall calculate the
expression for this classical cross section. One may wonder if there
is a connection between the quantum and classical regimes for finite
solenoid radius.  We find that the differential cross section
obtained from the first order QED calculation does not reduce to the
classical value in the $\hbar \to 0$ limit.  Surprisingly enough, it
is shown that the classical result can not be recuperated even if
higher order corrections are included.  To recover the classic
correspondence of the quantum scattering problem a suitable
non-perturbative methodology should be applied.

\section{Classical Cross Section}  
\label{sec:Classical_Result}  

Let us first consider the classical differential cross section of
charged particles by the magnetic field of a long solenoid of finite
radius $R$ and fixed magnetic flux $\Phi$.  Utilizing the classical
equation of motion the scattering angle as a function of the impact
parameter $b$ is obtained as
\begin{equation}
\theta(b) =  2 \arctan{ \left( \frac{ \sqrt{R^2-b^2} }{b+r_L} \right) },
\label{eq:theta(b)}
\end{equation}
where $r_L = {p c}/{e B}$ is the Larmor radius.
The impact parameter $b(\theta)$ is a multiple-valued function of
$\theta$; hence, the differential cross section requires to add the
two branches of the function, the result is worked out as
\begin{eqnarray}
\frac{1}{R} \frac{d\sigma_{\rho_L}(\theta)}{d\theta}
                       &=&   \left|{ \frac{\sin{\theta}}{2}
                                  \left({ \rho_L 
                                    + \frac{1 + \rho_L^2 \cos{\theta}}
                                           {2 \cos{(\theta/2)} \sqrt{ 1 - \rho_L^2\sin^2{(\theta/2)}}}
                                  }\right) 
                              }\right| + \nonumber \\
                        & &   \left|{ \frac{\sin{\theta}}{2}
                                  \left({ \rho_L 
                                    - \frac{1 + \rho_L^2 \cos{\theta}}
                                           {2 \cos{(\theta/2)} \sqrt{ 1 - \rho_L^2\sin^2{(\theta/2)}}}
                                  }\right) 
                              }\right| \Theta(|\rho_L| -1), 
\label{eq:secc_dif_clasica}
\end{eqnarray} 
where we defined a dimensionless parameter $\rho_L = r_L/R = p
R/2\beta$, with $\beta = e\Phi/2 \pi c $; and $\Theta(x)$ is the
Heaviside step function. Notice that in the low energy regime ($\rho_L
< 1$) the scattering angle covers all the range $\theta \in [- \pi
,\pi)$.  Instead for $\rho_L > 1$ there is a maximum allowed
scattering angle $\theta \in [0,\theta_{\mbox{max}}]$; where $
\sin(\theta_{\mbox{max}}) = 1/\rho_L. $

As expected, the Lorentz's force produces in general a classical DCS
that is not symmetric with respect to the forward direction ($\theta
=0$).  Furthermore, it is worthwhile to observe the highly nonlinear
dependence of the DCS on the coupling $\beta = e\Phi / 2 \pi c$.

The impenetrable limit $\rho_L \rightarrow 0 $ is obtained with $p R
\rightarrow 0$ and fixed $\Phi$; or considering the limit $\Phi
\rightarrow \infty $ with fixed $p R$; in both case the DCS reduces to
\begin{equation}
 \left.{\frac{d\sigma}{d\theta}}\right|_{\rho_L \rightarrow 0} = 
        \frac{R}{2}\left|{\sin(\theta/2)}\right|,
\label{eq:secc_dif_clasica-zero_radius_limit}
\end{equation}
a result that, as expected, is symmetric with respect to the forward
direction and independent of the coupling to the magnetic field.

Another interesting limit is obtained for high energy incident
particles with fixed magnetic flux: $pR \rightarrow \infty$ $(\rho_L
\gg 1)$. The scattered electrons are confined inside a narrow cone
aligned along the forward direction, defined by the maximum angle
$\theta_{\mbox{max}} \approx 1/ \rho_L$.  It is possible to show from
equation~(\ref{eq:secc_dif_clasica}) that the cross section reduces to
\begin{equation}
 \left.{ \frac{d\sigma}{d\theta} }\right|_{\rho_L \gg 1}  \approx
   R \, \theta \frac{1+\rho_L^2}{\sqrt{4-\rho_L^2\theta^2}}, 
  \hskip1.5cm  \vert \theta \vert  \le \theta_{\mbox{max}} . 
\label{eq:secc_dif_clasica_small_rho}
\end{equation}
We notice again the nonlinear dependence of the DCS on the coupling
$e\Phi$, a result that anticipates the incompatibility of the
classical result with the one that will be obtained in a quantum
perturbative calculation to any given finite order.

\section{Perturbative Quantum Analysis}
\label{sec:Quantum_Result}  

We now turn our attention to the calculation of the DCS in the quantum
regime.  The electron interacts with the gauge potential, that for the
finite radius solenoid can be represented as
\begin{equation}
 A_i=- \frac{\Phi}{2\pi}\epsilon_{ij3}x_j
        \begin{cases}
           \frac{1}{R^2}        & \text{for $r<R$} \\
           \frac{1}{x^2_1+x^2_2} & \text{for $r>R$}.
        \end{cases}
\label{ec:PotencialSolenoide}
\end{equation}
The interaction of the electron with the external magnetic field is
taken into account by introducing a dimensionless coupling factor $e
\Phi / \hbar c $ for each interaction of the electron with the
external field, and a factor related to the Fourier transformation of
the gauge potential $A_i$:

\begin{equation}
 - 2 i \frac{\hbar^2}{R} J_1(q_\perp R/\hbar) \epsilon_{ij3} 
       \frac{q_j}{q_\perp^3},
 \label{labelponer}
\end{equation}
where $q_\perp$ refers to the momentum perpendicular to the direction
of the magnetic field, and $J_1$ is the Bessel functions of first
kind.

The DCS was calculated in reference~\cite{Murguia_2003} to the lowest
perturbative order in $\beta = e\Phi/2 \pi c $, using free particle
incident and final asymptotic states, yielding
\begin{equation}   
\frac{d\sigma}{d\theta} =   
       \hbar \left({\frac{e\Phi}{Rc}}\right)^2   
       \frac{{\left| J_1(2\frac{p}{\hbar}R  
                     {\left|\sin{(\theta/2)}\right|})  
              \right|}^2}  
            {8\pi p^3 \sin^4{(\theta/2)}}.
\label{eq:dsigma_MM}   
\end{equation}
The previous result has the same form whether or not the final
polarization of the beam is actually measured. As can be observed, the
cross section is symmetric in the scattering angle $\theta$ with
respect to the forward direction.

The marked different behavior between the classical and quantum DCS
becomes evident; first from the symmetric behavior of the quantum
result, equation~(\ref{eq:dsigma_MM}), as compared to the asymmetric
structure of the classical one, equation~(\ref{eq:secc_dif_clasica}).
Furthermore, notice that the total quantum cross section is infinite,
in contrast to the finite value of $2R$ obtained for the classical
case.
More important is the fact that the quantum DCS in
equation~(\ref{eq:dsigma_MM}) is directly proportional to the coupling
$e\Phi$, while the classical DCS diverges as $e\Phi \rightarrow 0 $.

In order to consider the classical limit of the DCS in
equation~(\ref{eq:dsigma_MM}), we recall that according to Berry and
Mount~\cite{Berry_1972} and Gutzwiller~\cite{Gutzwiller}, the
implementation of the classical limit requires to look at the
situation in which the action quantities that appear in the
corresponding classical problem are considered as very large as
compared to $\hbar$~\cite{Note_References}.  Here we identify two
action variables, selected as: $p R$ and $ e\Phi / c$. It is then
convenient to define the dimensionless parameters $s_p = pR/\hbar$ and
$s_\Phi = e\Phi/\hbar c$. In term of these dimensionless parameters
the DCS in equation~(\ref{eq:dsigma_MM}) can be recast as
\begin{equation}
\frac{d\sigma}{d\theta} =   
       \frac{R}{8\pi} \frac{s_\Phi^2}{s_p^3}
       \left|{ \frac{J_1(2 s_p {\left|\sin{(\theta/2)}\right|})}
                    {\sin^2{(\theta/2)}} 
      }\right|^2.
\label{eq:DCS_MM_s_Phi-s_p}
\end{equation}
The classical limit is enforced by considering both $s_p \gg 1 $ and
$s_\Phi \gg 1 $.  We observe that the classical limit of the DCS in
equation~(\ref{eq:DCS_MM_s_Phi-s_p}) vanishes because it behaves as
$s_\Phi^2/s_p^4 \propto \hbar^2 \rightarrow 0$.  This result
establishes that the classical DCS can not be recovered in the
``classical limit'' of the quantum DCS calculated to first order in
$\beta = e\Phi/2 \pi c $.

\begin{center}
\begin{figure}[ht]
\includegraphics[width=0.7\linewidth]{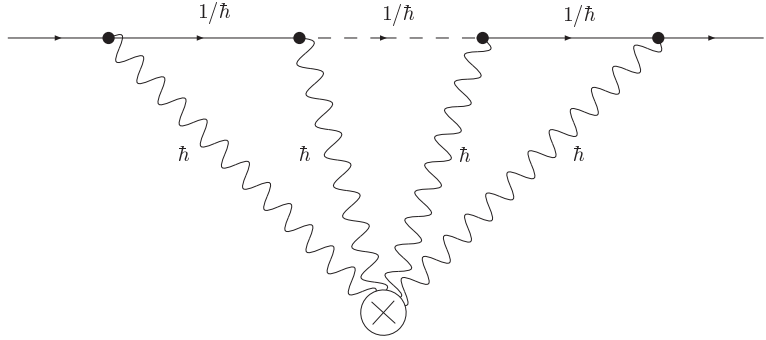}
 \caption{Feynman diagram and $\hbar$ power counting for an arbitrary
          order in $\beta = e\Phi/2\pi c$ of the scattering matrix for
          a solenoidal magnetic field. The wiggled lines
          represent the interaction with the
          external magnetic field while the straight lines represent the
          free-fermion propagators.}
  \label{figure:Feynman_diagram_all_orders}
\end{figure}
\end{center}

Higher order processes can be calculated using the Feynman rules for
the electron-solenoid scattering ~\cite{Murguia_2005}. Counting the
$\hbar$ power contributions to higher order diagrams (as the one
depicted in figure~\ref{figure:Feynman_diagram_all_orders}) and
assuming free particle asymptotic states, it can be shown that higher
orders in $\beta$ do not modify the leading $\hbar$ power contribution
to the scattering matrix; in fact higher order corrections in $\beta$
contribute with terms proportional to positive higher powers in
$\hbar$.

We recall that usual radiative corrections (higher powers in $\alpha$)
will in general contribute with positive $\hbar$ powers to the matrix
elements, hence they are not expected to be relevant in the classical
limit.  Consequently, for arbitrary finite order the perturbative
expansion in both $\beta$ and $\alpha$ produces a contribution
proportional to powers of $\hbar$, that cancels in the classical limit
of this process. Consequently the classical expression for the DCS can
not be recovered.

\begin{center}
\begin{figure}[ht]
\includegraphics[width=\linewidth]{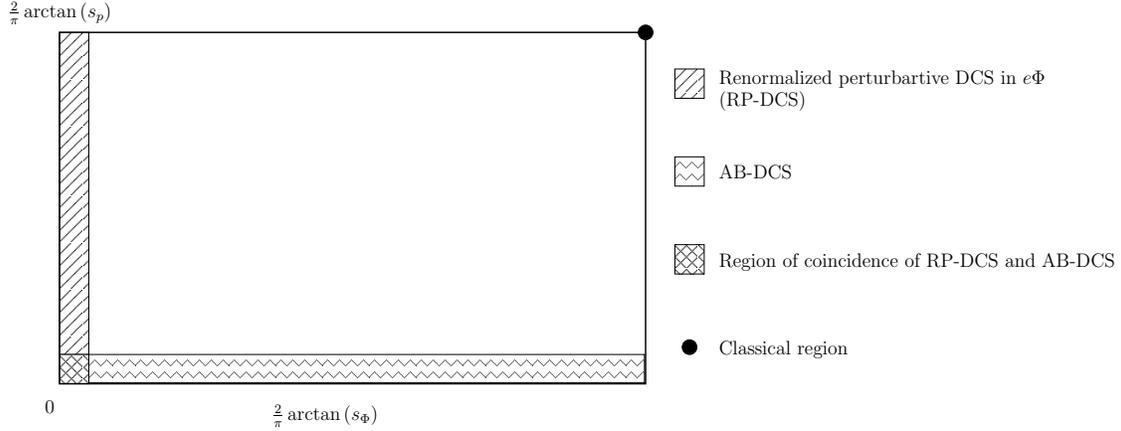}
 \caption{Diagram $s_p$ vs $s_\Phi$ for the quantum cross section of
          the scattering by a solenoidal magnetic field. The results for small
          $s_p$ and $s_\Phi$ are shown by the dashed regions. The classical
          region is represented by the dot in the upper right corner.}
 \label{figure:diagram_sP-sPhi}
\end{figure}
\end{center}

The various regions for the scattering electron-solenoid process are
schematically displayed in the diagram of
figure~\ref{figure:diagram_sP-sPhi}.
For illustrative purposes, the arc-tangent of $s_p$ and $s_\Phi$ are
normalized to unity.
There are depicted the regions in which
equations~(\ref{eq:secc_dif_AB}) and~(\ref{eq:dsigma_MM}) are valid,
including the renormalized perturbative terms in $\beta = e\Phi/2 \pi
c $.  Notice that the Aharonov-Bohm DCS is valid for small $s_p$;
whereas the perturbative results in $\beta $ are valid in the small
$s_\Phi$ region.  Both results coincide in the $s_p \rightarrow 0$ and
$s_\Phi \rightarrow 0$ region ~\cite{Murguia_2003}. It is expected
that the exact quantum calculation (valid for all values of $s_p$ and
$s_\Phi$) has the correct classical limit in the $s_p \rightarrow
\infty$ and $s_\Phi \rightarrow \infty$ region, which is depicted with
a dot in the upper right corner of the diagram.

\section{Conclusions}  
\label{sec:Conclusions}

In this paper we have studied the classical and quantum scattering of
an electron by a magnetic field confined in an infinite solenoid of
finite radius.  In the classical scenario the DCS shows a nonlinear
dependence on the coupling parameter $\beta = e\Phi/2 \pi c $ and a
general asymmetric behavior with respect to the forward direction.
The DCS obtained in the perturbative quantum regime displays marked
differences as compared with the classical one. The classical limit of
a corresponding quantum observable is characterized as the limit in
which all the relevant action quantities are considered very large as
compared with $\hbar$. We found that the classical DCS is not
recovered from the quantum DCS, even if higher order corrections are
included. We conclude that in general perturbative calculations easily
could drive to unappropriated results in the classical limit, because
in the perturbative regime typically at least one parameter remains
small in comparison with $\hbar$.
 
To recover the classical correspondence of the quantum scattering
problem a suitable non-perturbative methodology should be applied.  It
has been shown in~\cite{Murguia-Moreno-Torres} that an exact
expression for the quantum non-relativistic DCS can be obtained. Then
a combination of the large $\text{action\_variable}/\hbar$ limit, with
an stationary phase approximation for the evaluation of the partial
wave summation can be successfully implemented in order to correctly
derive the classical limit .

\bibliographystyle{aipproc}

\begin{thebibliography}{7}
\bibitem[Aharonov and Bohm(1959)]{AB}
Y.~Aharonov, and D.~Bohm, \emph{Phys. Rev.} \textbf{115}, 485 (1959).

\bibitem[Murguia and Moreno(2003)]{Murguia_2003}
G.~Murguía, and M.~Moreno, \emph{J. Phys.} \textbf{A36}, 2545 (2003).

\bibitem[Berry and Mount(1972)]{Berry_1972}
M.~V. Berry, and K.~E. Mount, \emph{Rept. Prog. Phys.} \textbf{35}, 315 (1972).

\bibitem[Gutzwiller(1990)]{Gutzwiller}
M.~C. Gutzwiller, \emph{Chaos in Classical and Quantum Mechanics}, Springer,
  New York, 1990.

\bibitem[Note()]{Note_References}
 We want to stress that both references~\cite{Berry_1972, Gutzwiller}
  present a general analysis of the classical limit and many references to
  relevant works on this subject are given there.

\bibitem[Murguía(2005)]{Murguia_2005}
G.~Murguía, \emph{Límite Clásico de la Dispersión Magneto-solenoidal de
  Partículas Cargadas}, Ph.D. thesis, Universidad Nacional Autónoma de México
  (2005).

\bibitem[Murguía et~al.(2008)]{Murguia-Moreno-Torres}
G.~Murguía, M.~Moreno, and M.~Torres, \emph{To be published}  (2008).

\end{thebibliography}

\end{document}